\documentclass[12pt,preprint]{aastex}

\newcommand{\myemail}{yuanqirong@njnu.edu.cn}
\newcommand{\Ha}{H$\alpha$}
\newcommand{\Msolar}{\mbox{\,$\rm M_{\odot}$}}

\slugcomment{accepted by the AJ}

\shorttitle{ Star Formation Properties of the Galaxies in A2255 }

\shortauthors{Yuan et al.}

\begin{document}

\title{MORPHOLOGICAL DEPENDENCE OF
STAR FORMATION PROPERTIES FOR THE GALAXIES IN THE MERGING GALAXY
CLUSTER A2255}


\author{\large Qirong Yuan\altaffilmark{1,2},
 Lifang Zhao \altaffilmark{1},
 Yanbin Yang \altaffilmark{2},
 Zhonglue Wen \altaffilmark{2},
 and Xu Zhou \altaffilmark{2}
}

\altaffiltext{1}{Department of Physics, Nanjing Normal University,
                 NingHai Road 122, Nanjing 210097, China; \\
                 \myemail}

\altaffiltext{2}{National Astronomical Observatories, Chinese Academy
of Sciences, Beijing 100012, China}

\begin{abstract}

The merging cluster of galaxies A2255 is covered by the Sloan
Digital Sky Survey (SDSS) survey. The physical parameters of 184
bright member galaxies derived from the SDSS data analyses by
Brinchmann et al. (2004b) allow a detailed study on the star formation
properties of the galaxies within a merging cluster at intermediate
redshift. In this paper we perform a morphological classification 
on the basis of the SDSS imaging and spectral data, and investigate 
the morphological dependence of the star formation rates (SFRs) 
for these member galaxies. As we expect, a tight correlation between 
the normalized SFR by stellar mass (SFR/M$_*$) and the \Ha\ equivalent 
width is found for the late-type galaxies in A2255. The correlation 
of SFR/M$_*$ with the continuum break strength at 4000 \AA\ is 
also confirmed. The SFR/M$_*$ -
M$_*$ correlation is found for both the early- and late-type galaxies,
indicating that the star formation activity tends to be suppressed
when the assembled stellar mass (M$_*$) increases, and this correlation
is tighter and steeper for the late-type cluster galaxies. Compared
with the mass range of field spiral galaxies, only two massive late-type 
galaxies with M$_*>10^{11}$\Msolar\ are survived in A2255, suggesting that
the gas disks of massive spiral galaxies could have been tidally
stripped during cluster formation. Additionally, the SFR variation with
the projected radial distance are found to be heavily dependent upon
galaxy morphology: the early-type galaxies have
a very weak inner decrease in SFR/M$_*$, 
while the inner late-type galaxies tend to have higher SFR/M$_*$ values 
than the outer late-types. This may suggest that the galaxy-scale 
turbulence stimulated by the merging of subclusters might have 
played different roles on early- and late-type galaxies, which
leads to a suppression of the star formation activity for E/S0
galaxies and a SFR enhancement for spiral and irregular galaxies.

\end{abstract}



\clearpage

\section{INTRODUCTION}

To study the star formation history of the universe is one of the
major tasks of extragalactic astronomy. The current star formation
rates (SFRs) of galaxies and the variation with morphology, environment
and some physical properties of galaxies are crucial in our
understanding of the evolution of galaxies (Kennicutt 1998;
Brinchmann et al. 2004a). Some studies show that the galaxy-scale 
turbulence plays a key role on star formation of galaxies (e.g.
MacLow \& Klessen 2004; and references therein). However, the origin
and consequence of this turbulence are still unclear, it is
reasonable to expect that the star formation activity should be a
function of many variables. A large number of investigations focus on
finding the possible correlations of the SFRs
of galaxies with some global quantities, such as stellar mass
(Kauffman et al. 2003a,b), morphological classification (Kennicutt,
Tamblyn, \& Congdon 1993; Kennicutt 1983; Kennicutt 1998),
metallicity (Brinchmann et al. 2004a; Tremonti et al. 2004),
luminosity (e.g., Bell 2003), and redshift (Butcher \& Oemler 1984;
Finn, Zaritsky, \& McCarthy 2003).

It is widely accepted that above-mentioned physical properties of
galaxies are fundamentally linked with their gravitational environment
(e.g., Dressler 1980; Butcher \& Oemler 1984). Of great concern is
whether the star formation properties of cluster galaxies are similar
to those of field galaxies. With the traditional H$\alpha$
spectroscopy (Kennicutt 1983), the star formation rates (SFRs) for
nearby galaxies ($z<0.3$) are estimated by many investigators. The main
advantages of this approach are direct measurements of ionizing flux
from young stars and insensitivity to extinction than the [\ion{O}{2}]
measurements. Based on the \Ha\ data, Kennicutt et al. (1983,
1984) study four nearby clusters, and they find that the cluster and
field spirals with the same morphological type have similar SFRs in
three of four clusters. However, the difference between field and
cluster spiral galaxies is found by Moss \& Whittle (1993), and such
difference is likely to be dependent upon morphology (Moss \& Whittle
2000).

The SFRs in late-type galaxies are found to be closely related to
their gas content (Kennicutt 1998). To look at which environmental
processes have actually affected the gas content of the spiral galaxies
within a nearby cluster, Koopmann et al. (2004) compare the \Ha\
morphologies of the spiral galaxies in the Virgo cluster with those of
isolated field spirals, and find obout half of the Virgo spiral
galaxies having truncated \Ha\ disks. Considering that the truncated
\Ha\ disks are relatively rare in isolated spiral galaxies, this 
result suggests that many Virgo spiral galaxies could have experienced 
intracluster medium (ICM) - interstellar medium (ISM) stripping and 
significant tidal effects.
 Gavazzi et al. (2002) investigate the star formation properties
as a function of the clustercentric projected distance for 369
late-type galaxies in the Virgo, Coma and A1367 clusters, and they
find that the bright spirals in the Virgo cluster tend to decrease 
their SFRs inward, and the fainter late-types show no or reverse trend.
Additionally, convincing evidence of the star formation suppression
of the cluster galaxies has been found in some rich clusters with higher
redshifts (Couch et al. 2001; Balogh \& Morris 2000; Balogh et al.
2002; Finn et al. 2004).

This paper explores the properties of star formation for the merging
cluster of galaxies A2255 that is covered by the Sloan
Digital Sky Survey (SDSS) and the Beijing-Arizona-Taiwan-Connecticut
(BATC) multicolor photometric survey. There are 132 early-type 
galaxies and 52 late-types in this cluster with their absolute 
magnitudes in the SDSS $r$ band ($M_r$) brighter than $-20.0$, 
which provides a good sample for
studying the SFR variation of cluster galaxies with
morphology and projected radial distance. Another main motivation
of this study is to find possible links among the physical properties
of cluster galaxies. For a better understanding of the environmental
effects of the star formation properties of galaxies
at different evolutional stages, it is illuminating to compare our 
results with those of field galaxies and the high-$z$ cluster
galaxies.

We start in \S 2 with describing the relevant physical parameters for
the galaxies in A2255. The star formation properties for these
cluster galaxies, as a function of other physical parameters of
galaxies, are shown in \S 3. We then compare the properties with nearby
field galaxies and high-$z$ cluster galaxies, and discuss their
implications in \S 4. Finally, the conclusions are given in \S5. A
Friedmann-Robertson-Walker cosmology with $(\Omega_m,\Omega_\Lambda)
=(0.3,0.7)$ and $H_0=70$ km/s/Mpc is assumed throughout this paper.

\section{Physical parameters for cluster galaxies}

A2255 is a cluster of galaxies with an intermediate redshift of
$z=0.0806$ (Struble \& Rood 1999, giving a distance modulus of 37.8),
and richness class 2 (Abell 1958). Using the data from the the
SDSS Data Release 2 (DR2) and the BATC multicolor photometric survey,
we have explored the color-magnitude relation, luminosity function
and dynamics of the member galaxies, and a direct dynamical evidence
for an on-going merger has been found for this cluster
(Yuan et al. 2003).

The SDSS aims at targeting $10^6$ galaxies with $r$-band 
magnitude $r<17.77$ for spectroscopy (Strauss et al. 2002).  Based
on a large sample of spectroscopically-confirmed galaxies provided
by the SDSS DR2 (Abazajian et al. 2004) and the
new models of stellar population synthesis (Bruzual \& Charlot 2003;
Charlot et al. 2002; Charlot \& Longhetti 2001), Brinchmann et al.
(2004b) have derived and compiled the physical quantities concerning
the star-formation properties for more than $2 \times 10^5$
galaxies, including stellar mass, current SFRs, gas-phase metallicity,
absorption line indices, the emission line measurements, and so on.

The parameters relevant to this work are estimated mainly by
following investigations: (i) Kauffmann et al. (2003a,b) develop a
method to constrain the star formation history, dust attenuation and
stellar mass ($M_*$) of the specified galaxy on the basis on the
continuum break at 4000 \AA\ ($D_n$) and the $H\delta$ absorption
line. The strength of continuum break at 4000 \AA\ is a prominent
feature in galaxy spectra resulted from a large number of metal
lines. This can be treated as a powerful age estimator since the
young and hot stellar population show a very weak 4000\AA\ break. The
strength of continuum break is also defined in Tremonti (2002).
(ii) Brinchmann et al. (2004a) make use of the methodology
in Charlot et al. (2002) and the emission line models in Charlot \&
Longhetti (2001) to derive the SFRs inside the fibre aperture. Then,
the total SFRs in galaxies are derived by the aperture correction.

We select all the SDSS galaxies with $14.5 < r < 17.77$ 
in the BATC-targeting sky region of 59 $\times$ 59 arcmin$^2$,
defined by a right ascension range from $17^{\rm h}08^{\rm m}03^{\rm
s}.7$ to $17^{\rm h}16^{\rm m} 58^{\rm s}.6$, and a declination range
from $63^{\circ}36'31''.5$ to $64^{\circ}35'28''.8$ (J2000.0) (Yuan et
al. 2003). As a result, we obtain 184 spectroscopically-confirmed
member galaxies with $0.070 < z < 0.097$, of which the physical
properties are publicly released  at
http://www.mpa-garching.mpg.de/SDSS/ by Brinchmann et al. (2004b). The
completeness of this sample reaches at least 90\% down to a $r$-band 
absolute magnitude ($M_r$) of $\sim -20.0$. This should be an ideal
sample for investigating the star formation properties of the
galaxies within a merging cluster at intermediate redshift.

\section{Results}

\subsection{Morphological classification}

In order to classify the cluster galaxies into early- and late-type
galaxies, we have collected their SDSS imaging and spectral data, 
as well as some photometric parameters relevant to the morphology, such as the
likelihoods for radial profile fitting with the de Vaucouleurs (1948)
$r^{1/4}$ model ($L_{\rm deV}$) and the exponential model ($L_{\rm
exp}$), the concentration index characterized by the ratio between two
galaxy radii which contain 90\% and 50\% of Petrosian flux respectively
($R_{p,90}/R_{p,50}$), the color index ($u-r$), and the continuum break 
strength at 4000\AA\ ($D_n$). 

Though the concentration index and fitting likelihoods parameterize the 
light profiles of galaxies, the eye-ball inspecting should be the most 
straightforward way for finding some morphological characteristics (e.g., 
spiral arms, bars, rings, dust lane, etc.). For a secure morphological
classification, it is important to directly inspect the SDSS images of 
galaxies. For about half of galaxies in A2255, the eye-ball
classification is quite confident. It can be understood that the regular 
elliptical galaxies have $L_{\rm deV} \gg L_{\rm exp}$ and/or 
$R_{p,90}/R_{p,50} > 2.6$, while the disk-dominated spiral galaxies 
have $L_{\rm exp} \gg L_{\rm deV}$ and/or $R_{p,90}/R_{p,50} < 2.5$. 
For the galaxies with significant likelihood difference (i.e.,  
$|\ln L_{\rm deV}-\ln L_{\rm exp}|>150$) and one of profile-fitting 
likelihoods greater than $e^{-75}$ 
(i.e., $\ln\{L_{\rm deV} | L_{\rm exp}\}>-75$),
it is completely unambiguous to classify the galaxies with larger
$L_{\rm deV}$ as the early-types, whereas the late-types for the
galaxies with larger $L_{\rm exp}$. 

Nevertheless, the spatial resolution (0.396''/pixel) of SDSS images
is not high enough for a secure classification of some faint galaxies.
Furthermore, some late-type bulge-dominated (Sa-Sb) galaxies are likely 
to be misclassified as the early-type galaxies because their light
profiles cannot be distinguished from the early-types (Scodeggio et al.
2002). The template spectra of nearby normal galaxies (Kennicutt 1992)
show that the Sa-Sb spiral galaxies usually have week \Ha\ emission
feature which should be present in their SDSS spectra ($R=1800$).
Therefore, it is necessary to inspect the spectral feature for further
classification, particularly for the galaxies with low fitting 
likelihoods for both profile models. 
For ease to compare with the spectrophotometric atlas of nearby normal
and peculiar galaxies with known morphologies (Kennicutt 1992), we
transform the SDSS spectra to the rest frame defined by the given
redshifts, and then normalize the spectra by the flux at rest-frame
wavelength 5500\AA\ . For the majority of cluster galaxies, 
the spectral classification into the early- and late-types is 
unambiguous, and their images and relevant morphological parameters 
(i.e., $L_{\rm exp}$, $L_{\rm deV}$, and $R_{p,90}/R_{p,50}$) are 
consistent with each other. For small number of ambiguous cases in 
the S0/Sa separation, we classify the galaxies with the \Ha\ equivalent 
widths (EWs) greater than 1.0 \AA\ as the late-types in practice. 

Finally, We obtain 132 early-type galaxies and 52 late-type 
galaxies in A2255. It should be noted that there are 5 early-type 
galaxies with strong emission lines (i.e., EW(\Ha)$>$5\AA) in our sample. 
Their morphology-relevant parameters are given in Table 1. According to 
the diagnostics of the emission lines in spectra (Baldwin et al. 1981),
four of these five early-type galaxies are found to be the active 
galactic nuclei (AGNs), and one is a {\em star-forming} (SF) galaxy.


\begin{table}[ht]
\caption[]{A list of five early-type galaxies with strong emission lines (i.e.,EW(\Ha)$>$ 5\AA)}
\begin{tabular}{ccccccl}   \hline   \hline
\noalign{\smallskip}
Object Name & $r$ & $\ln L_{deV}$ & $\ln L_{exp}$ & \hspace{-4mm} $R_{p,90}/R_{p,50}$ & \hspace{-4mm}$D_n$ 
& Remark\\
\noalign{\smallskip}   \hline \noalign{\smallskip}

SDSS J171315.45+642418.3 & 17.0 &-10.744& -263.037 & 2.573 & 1.4729 
&  SF Galaxy\\
SDSS J171429.16+641656.3 & 16.6 &-71.789 & -369.118 & 2.367 & 1.4852 
&  BLAGN \\
SDSS J171221.14+643418.1 & 17.2 & -0.666 & -188.854 & 2.498 & 1.4749 
& NLAGN \\
SDSS J171302.13+634758.8 & 17.3 &  -0.038& -223.167 &  2.842 & 1.6238 
& NLAGN \\
SDSS J171052.51+633917.2 & 16.3 & -74.485& -1529.74& 2.872 & 1.6591 
& AGN(LINER) \\

\noalign{\smallskip}   \hline
\end{tabular}
\end{table}

\begin{figure}[tb]
\epsscale{0.9} \plotone{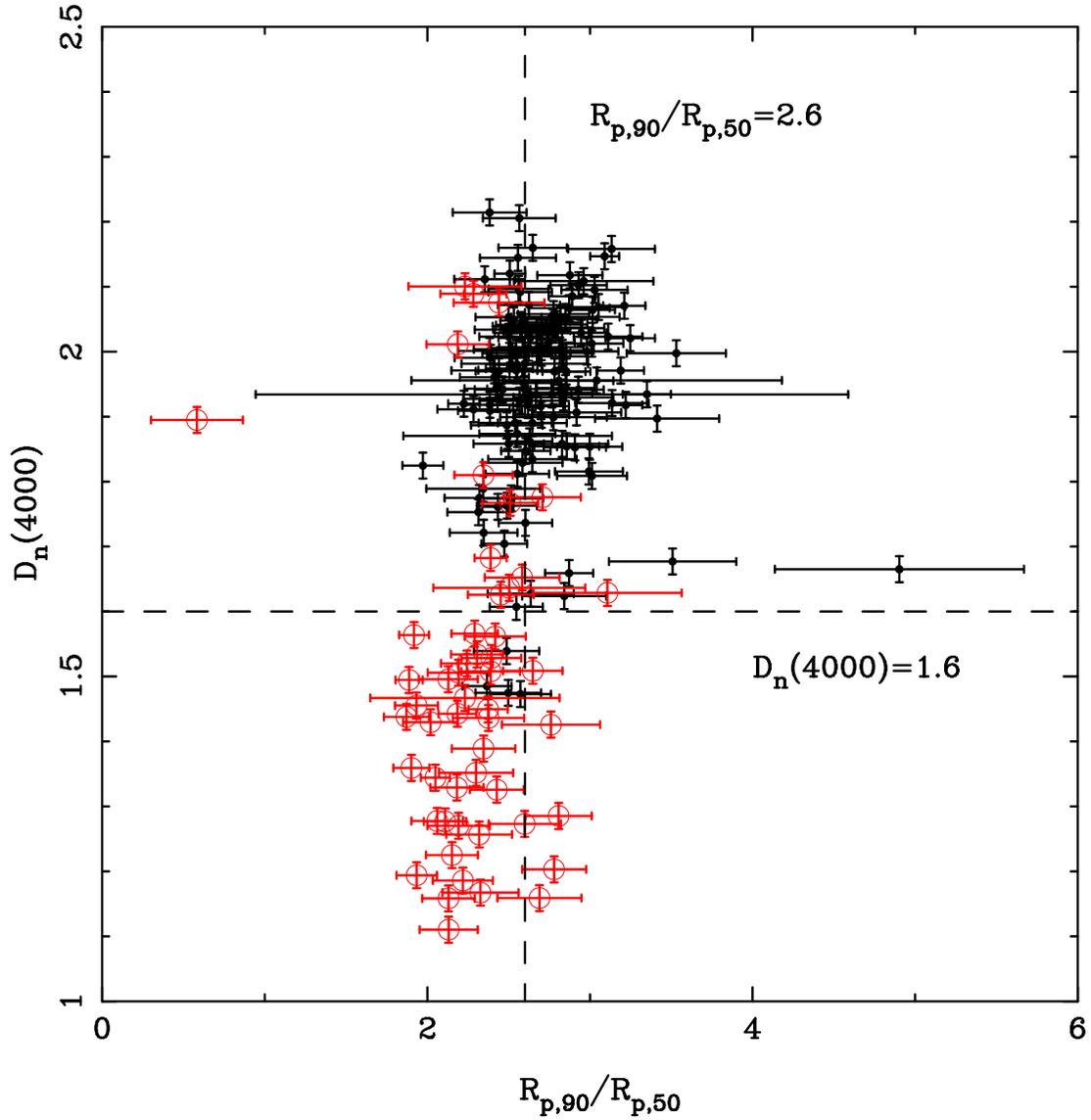} \caption {The continuum
break strengths ($D_n$) at 4000 \AA\ as a function of the radius
ratios ($R_{p,90}/R_{p,50}$) for 132 early-type galaxies
({\em solid points}) and 52 late-types ({\em open circles}) in
A2255.
\label{Dn_r90_50}}
\end{figure}

The criterion of $R_{p,90}/R_{p,50}>2.6$ is adopted for selecting the
elliptical galaxies by many investigators (e.g., Shimasaku et al. 2001;
Padmanabhan et al. 2004). In addition, Kauffmann et al. (2003a) use a
cut of $D_n > 1.6$ to pick up the elliptical galaxies. It is
interesting to check the efficiency of these criteria of morphology
classification. Fig.~\ref{Dn_r90_50} shows the continuum break strengths 
($D_n$) of the galaxies in A2255 as a function of the radius ratios
($R_{p,90}/R_{p,50}$). The early-type galaxies are denoted with
solid points, while the late-types with open circles. Two criteria
used in the literature are also marked in Fig.~\ref{Dn_r90_50}. 
It can be easily seen that $D_n$ is a better morphology indicator for 
the nearby {\em
cluster} galaxies. With the criterion $D_n > 1.6$, we can pick up 128
($\sim$ 97\% of 132) early galaxies, and 12 late-type galaxies are
misclassified as the early-types. However, if we simply adopt 
$R_{p,90}/R_{p,50}>2.6$ to construct a sample of the early-type galaxies 
in A2255, nearly 30\% of the early-types will be lost.

It should be pointed out that the $R_{p,90}/R_{p,50}-D_n$ relation is
heavily dependent upon gravitational environment. The slope at which 
$D_n$ increases with $R_{p,90}/R_{p,50}$ tends to be flat for
the galaxies in low-density regions (see figure 7 in Kauffmann
et al. 2004). Therefore, for selecting the {\em field} elliptical
galaxies, concentration parameter $R_{p,90}/R_{p,50}$ becomes
more effective than the continuum break strength $D_n$. Both criteria 
are taken by Shimasaku et al. (2001) to ensure their sampling of 
elliptical galaxies from the SDSS galaxies.

\subsection{The SFR tracers}

There are different indicators of galaxy SFR at different
wavelengths. In optical band, the traditional SFR calibration
uses the \Ha\ emission line luminosity. 
The [\ion{O}{2}] emission is also an important tool for estimating
current SFRs for the galaxies with $z>0.5$ (Kennicutt 1998), which is
based on the fact that there is a good correlation between observed
[\ion{O}{2}] line fluxes and observed \Ha\ fluxes (i.e., prior to any
obscuration corrections). Kennicutt (1992) sets up the SFR calibration with 
the [\ion{O}{2}] fluxes by using a sample of 90 nearby ($z<0.03$) 
galaxies, and obtains the empirical ratio  
EW([\ion{O}{2}])/EW(H$\alpha$+[\ion{N}{2}]) = 0.4.

\begin{figure}[tb]
\epsscale{0.9} \plotone{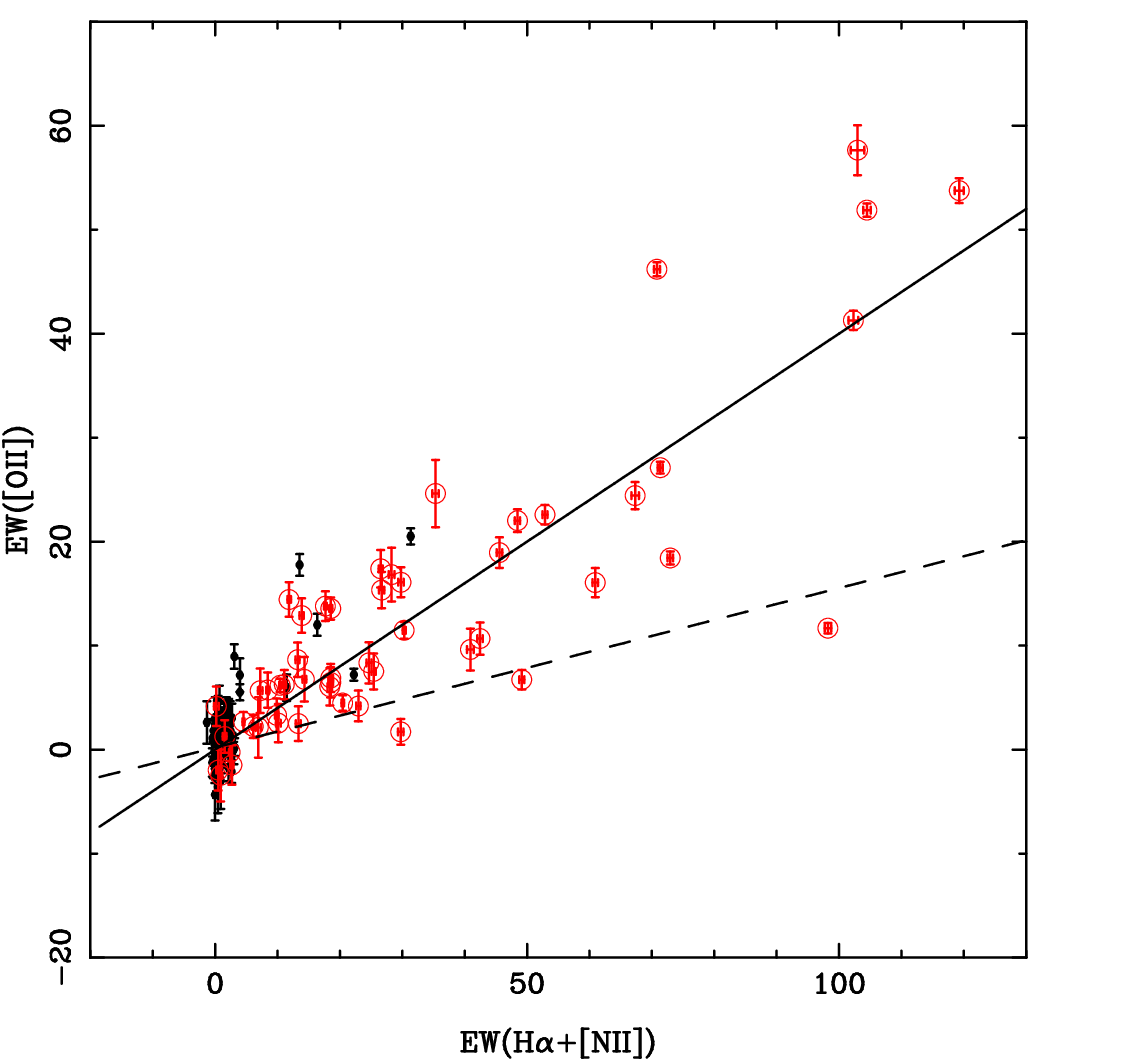} \caption {The correlation between
the equivalent widths of the composite line \Ha+[\ion{N}{2}] and the
[\ion{O}{2}] line. Denotion the same as in Fig.~\ref{Dn_r90_50}. 
We give the solid line corresponding to the observed ratio
EW([\ion{O}{2}])/EW(H$\alpha$+[\ion{N}{2}])=0.4 (from Kennicutt 1992),
and the dashed line representing the line flux ratio [\ion{O}{2}]/\Ha
=0.23 (from Hopkins et al. 2003), if we take the assumption of
[\ion{N}{2}]/H$\alpha$=0.5 (Kennicutt 1992).
\label{ew_o2_ha}}
\end{figure}

To avoid the additional uncertainties produced by the aperture 
corrections of the emission line fluxes, we sketchily observe the 
correlation between the equivalent widths (EWs) of these two lines.
For compatibility, we give the plot of the [\ion{O}{2}] EWs versus the
composite EWs of H$\alpha$+[\ion{N}{2}] for 184 member galaxies in 
Fig.~\ref{ew_o2_ha}. The Kennicutt's relation ({\em solid line} in 
Fig.~\ref{ew_o2_ha}) can be used to fit the EW correlation for 
184 A2255 galaxies very well. Jansen et al.
(2001) point out that the flux ratio [\ion{O}{2}]/H$\alpha$ should be
luminosity and metallicity dependent. Therefore, the appropriate flux
ratio should be influenced by the selection effects of various samples
of galaxies. The median flux ratio [\ion{O}{2}]/H$\alpha$ = 0.23 is
derived by Hopkins et al. (2003) with a sample of 752 SDSS
SF galaxies. Our result does not coincide with
this flux ratio, denoted with dashed line in Fig.~\ref{ew_o2_ha} when
[\ion{N}{2}]/H$\alpha$=0.5 (Kennicutt 1992) is assumed. This is probably
because our sample only contains the cluster galaxies predominated by
the early-type galaxies.

\subsection{The SFR variation with morphology}

\begin{figure}[tb]
\epsscale{0.9} \plotone{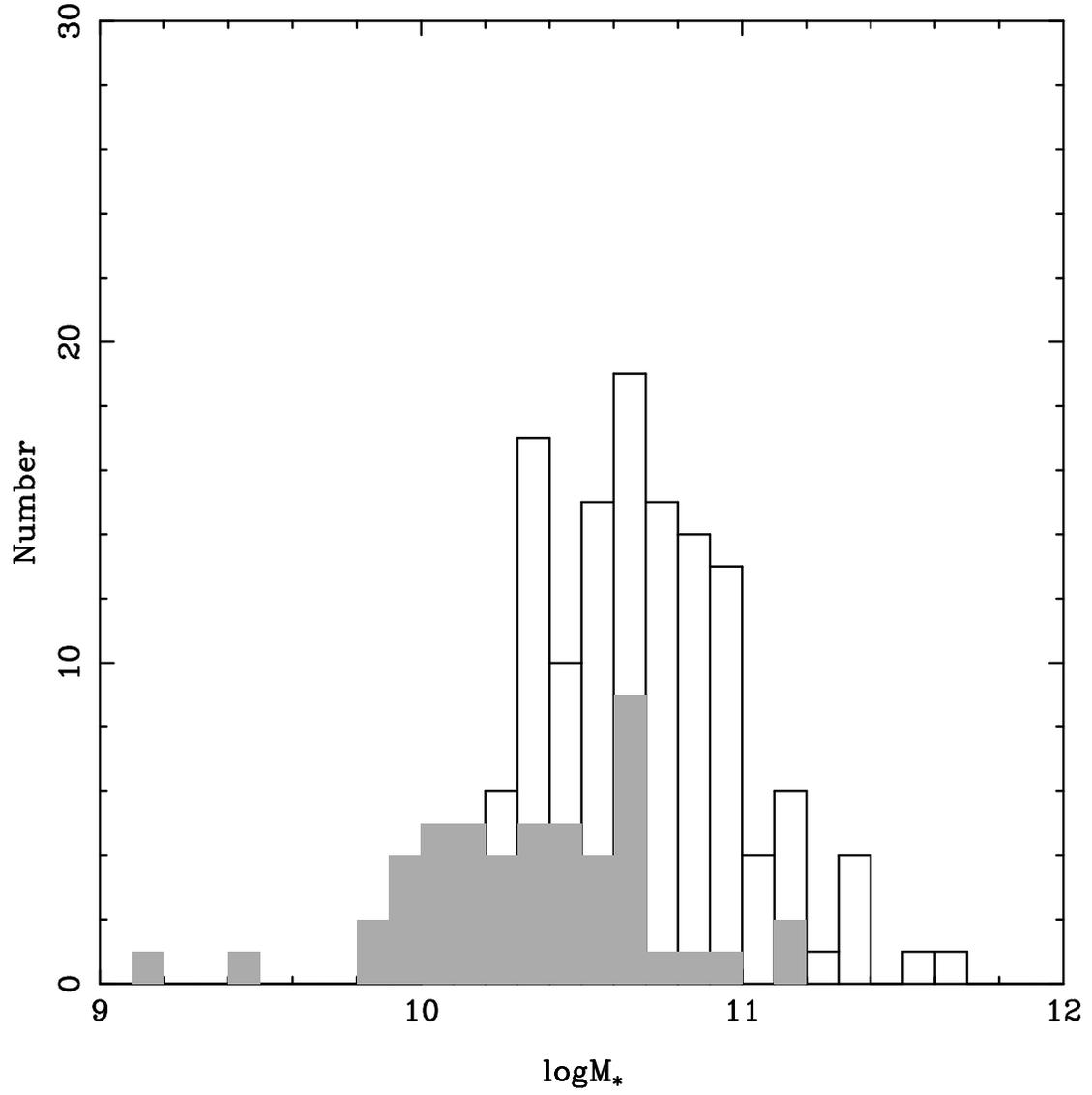}
\caption{Distribution of the stellar masses for 132 early-type galaxies
({\em line histogram}) and 52 late-type galaxies ({\em filled histogram}).
\label{n_m_star}}
\end{figure}

Fig.~\ref{n_m_star} shows the distributions of the stellar masses 
($M_*$) for the early-type galaxies ({\em open histogram}) and the late-types
({\em filled histogram}). The early-type galaxies occupy a broader
range (i.e., more than two orders of magnitude) in stellar mass, while
the late-type galaxies cover a range of about one order of magnitude.
We can find a significant trend towards old stellar populations for the
galaxies with larger stellar masses. The typical stellar mass of the
late-type galaxies is less by $\sim$ 0.4 dex than that of the
early-types. This histogram supports the finding by Kauffmann et al.
(2003b) that there is a rapidly increasing fraction of galaxies with
old stellar populations for the massive galaxies with $M_*>3\times 10^{10}$
\Msolar. In our sample there are only two late-type {\em cluster} galaxies
having their stellar masses larger than $10^{11}$ \Msolar.


\begin{figure}[tb]
\epsscale{0.9} \plotone{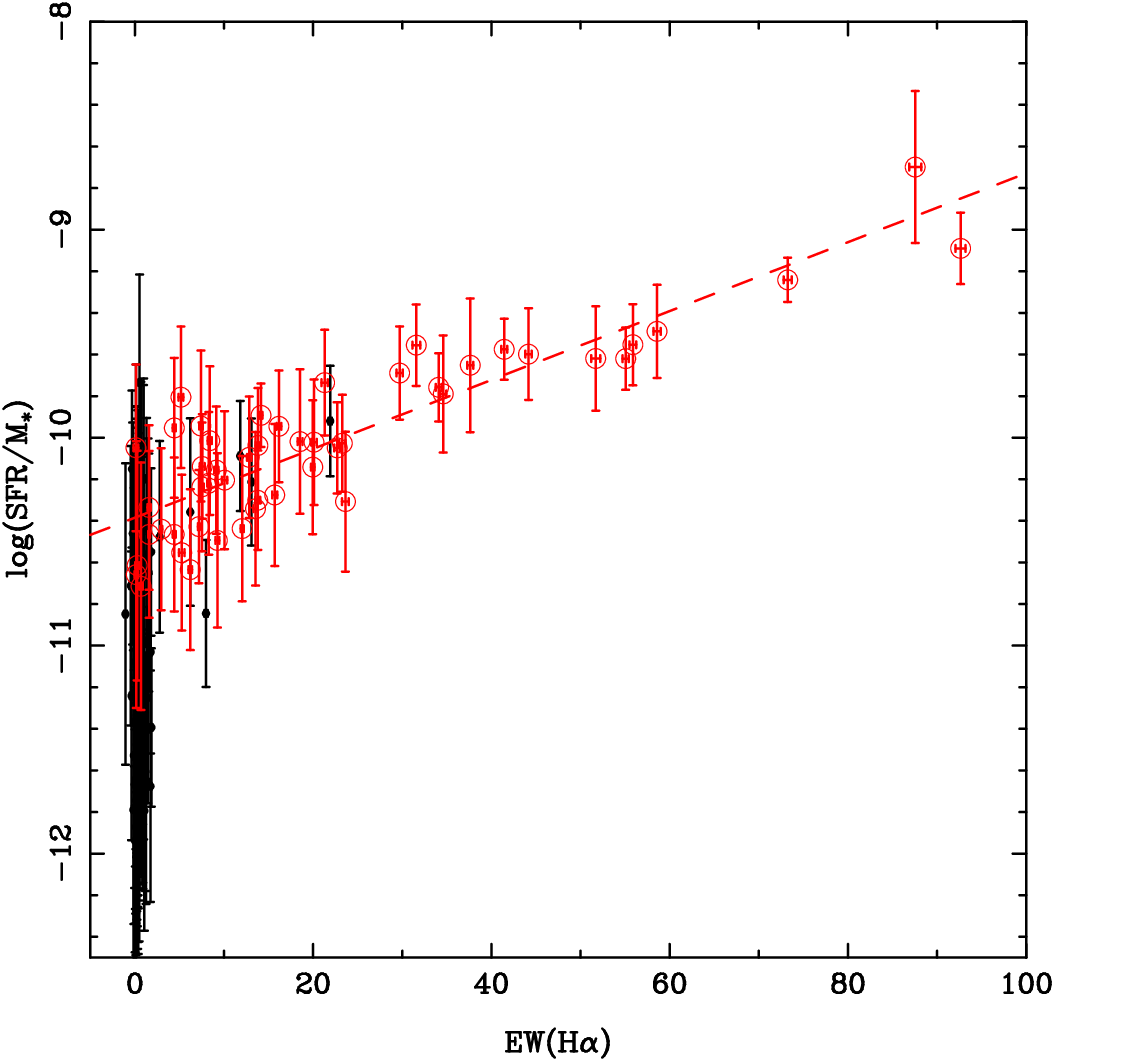} \caption
{The relation between the {\em specific SFRs} (SFR/$M_*$)
and the equivelent width of the H$\alpha$ emission line for 132
early-type galaxies ({\em solid points}) and 52 late-type galaxies 
({\em open circles}) in A2255. The dashed line represents a linear 
fit for the late-type galaxies only.
\label{sfr_m_ha}}
\end{figure}


The {\em specific SFR}, defined as SFR/$M_*$, is a key parameter for
measuring the rate that new stars add to the assembled mass of a
galaxy. It is common to use this normalized SFR by stellar mass to
investigate the relationship between the star formation activity and
the physical properties (e.g., Brinchmann et al. 1997). Brinchmann et
al. (2004a) find a strong correlation between the SFR and stellar mass
over a significant range in $M_*$/\Msolar\ from $3\times10^{6}$
to $10^{10}$ (see the figure 17 therein). Fig.~\ref{n_m_star} shows 
that a vast majority of the galaxies in A2255 have their stellar masses 
larger than $10^{10}$\Msolar, so the SFR-$M_*$ correlation breaks down 
for our sample.

Since the EW is defined as the emission-line luminosity normalized to
the adjacent continuum flux, the EW of \Ha\ emission line is also a 
direct measure of the SFR per unit red luminosity (Kennicutt 1998). 
It can be expected that there should be a tight correlation between the
\Ha\ EW and SFR/M$_*$, when a fixed mass-to-luminosity ratio is 
assumed for the late-type galaxies. This SFR/M$_*$-EW(\Ha) correlation 
is well shown in Fig.~\ref{sfr_m_ha}. A linear fitting is performed for 
the late-type 
galaxies with a wide range of the \Ha\ EWs, and we obtain
$\rm \log(SFR/M_*) = 0.017(\pm0.001)EW(H\alpha)-10.385(\pm0.041)$. This
correlation is so tight that its correlation coefficient reaches 0.877 
and the $rms$ dispersion in log(SFR/M$_*$) is 0.20. Meanwhile, this correlation 
demonstrate the reliability of the specific SFRs derived by Brinchmann
et al. (2004a) and Kauffmann et al. (2003a,b).

\begin{figure}[tb]
\epsscale{0.9} \plotone{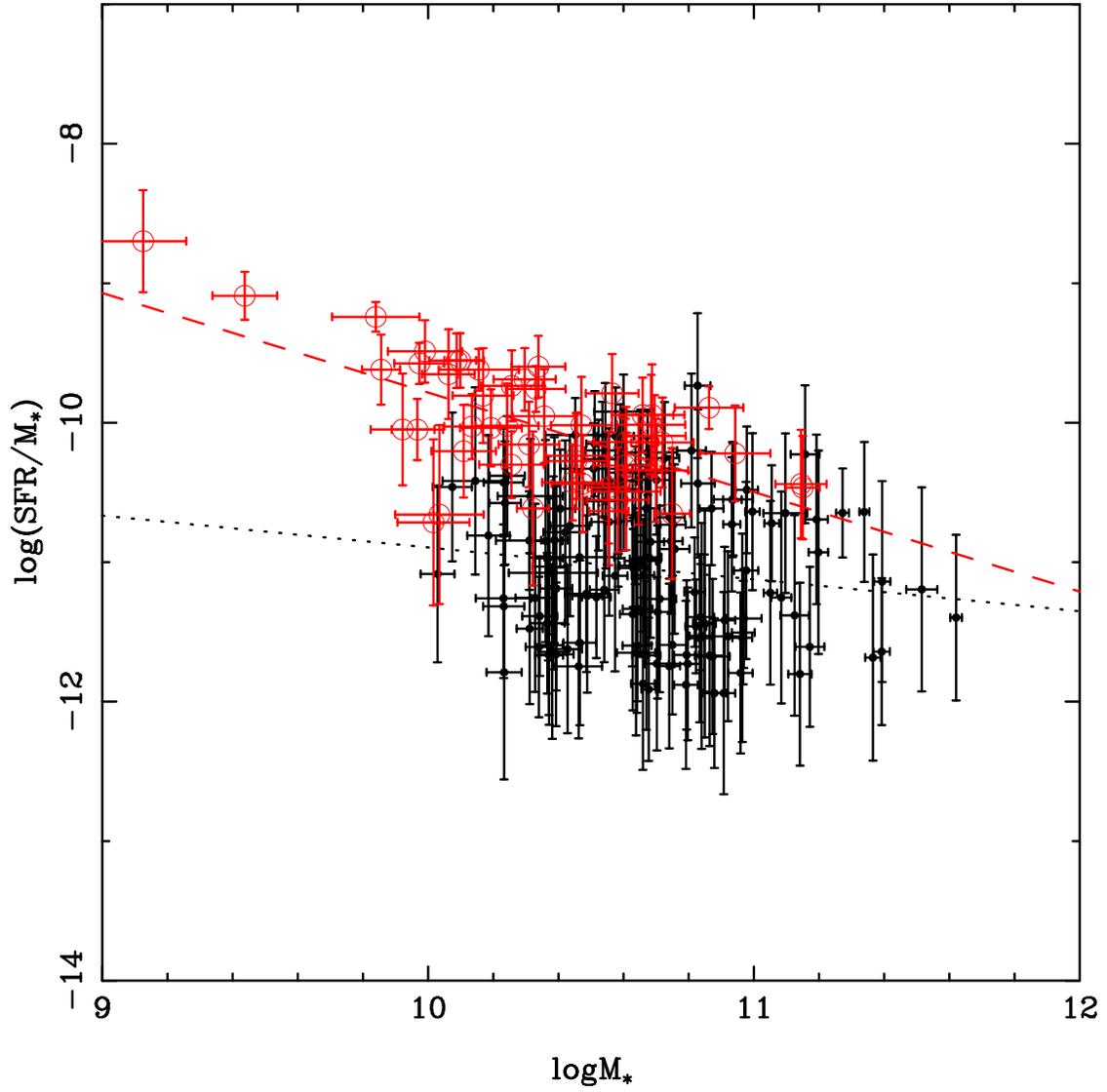}
\caption{The relation between the {\em specific SFRs} (SFR/$M_*$)
and the stellar masses ($M_*$) for 132 early-type galaxies
({\em solid points}) and 52 late-type galaxies ({\em open circles}) 
in A2255.
\label{sfr_m_m}}
\end{figure}

Fig.~\ref{sfr_m_m} shows the relation between the {\em specific} SFR 
and the stellar mass for the galaxies in A2255. The late-type galaxies 
tend to have higher SFR/$M_*$ values than
the early-types. Most of the early-type galaxies have lower fraction of
the newly-formed stellar masses, with respect to their large assembled
masses. The early-type
galaxies in A2255 cover a range from $10^{-12}$ to $10^{-10}$ in the
SFR/$M_*$ domain, and the specific SFRs decline slightly when their
stellar masses grow. The weak SFR/$M_*$-$M_*$ correlation can be
expressed by a linear fit of $\log(SFR/M_*)=
-0.23(\pm0.15)\,\log M_*-8.61(\pm1.59)$, which is shown in Fig.~\ref{sfr_m_m} 
with the dotted line. The correlation coefficient is only $-0.135$,
and the $rms$ dispersion of log(SFR/$M_*$) is 0.52. For the late-type
galaxies, however, one can see a clear trend that the star formation
activity decreases when the stellar mass increases (see the dashed
line), with a linear fit of
$\log(SFR/M_*)=-0.71(\pm0.12)\,\log M_*-2.65(\pm1.27)$. The correlation
coefficient is $-0.639$, and the $rms$ dispersion is 0.33. Keep in mind
that there is a significant difference between the slopes of
these two relations, and its implication will be addressed in \S 4.


\begin{figure}[tb]
\epsscale{0.9} \plotone{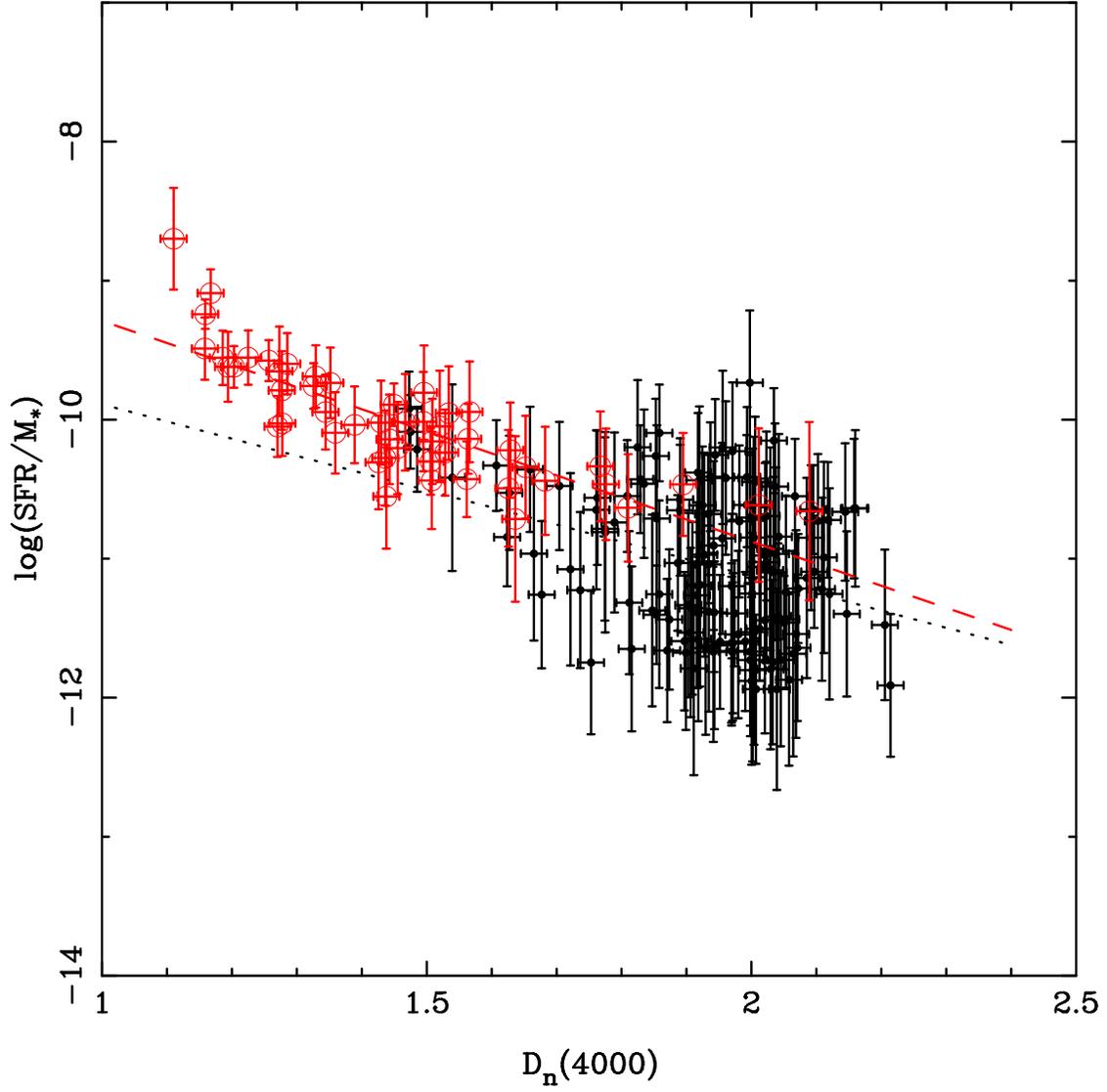}
\caption{The relationship between the
normalized SFR by stellar mass (SFR/M$_*$) and the continuum break
strength ($\rm D_n$) at 4000 \AA. Notation the same as in 
Fig.~\ref{Dn_r90_50}. 
\label{sfr_m_dn}}
\end{figure}

Brinchmann et al. (2004a) find a tight correlation between the
specific SFR (SFR/M$_*$) and the 4000\AA\ break strength ($D_n$) for
the SDSS SF galaxies with the signal-to-noise ratio $S/N > 3$ in the
emission lines \Ha, H$\beta$, [\ion{O}{3}] and [\ion{N}{2}]. Because
the SF galaxies with strong emission lines become rare in the rich
clusters, this correlation has been extended to the non-SF galaxies
for estimating the specific SFRs. We plot the relation between
SFR/M$_*$ and $D_n$ for the galaxies in A2255 in Fig.~\ref{sfr_m_dn}. 
A very tight correlation can be easily found for the late-type galaxies 
({\em open circles}) with smaller continuum break strengths. A linear 
fitting is given for the late-type galaxies, with an
expression of $\log(SFR/M_*)=-1.59(\pm0.16)\,\log M_*-7.70(\pm0.23)$
({\em dashed line}). The correlation coefficient is $-0.825$,
and the $rms$ dispersion is 0.24.

The SFR/M$_*$ - $D_n$ correlation becomes weak for the early-type
galaxies with a larger $rms$ dispersion of 0.49 and a similar slope of
$-1.24$($\pm0.30$). The continuum break strengths for the majority of
early-type galaxies are greater than 1.6. The distribution in the
$SFR/M_*$ - $D_n$ space for the early-type galaxies in A2255 is in
good accordance with the contour map shown in the figure 11 of Brinchmann
et al. (2004a), suggesting that the majority of the SDSS SF
galaxies with strong continuum break at 4000 \AA\ are the early-type
galaxies within the dense environment.

\subsection{The radial distribution of $\rm SFR/M_*$}

\begin{figure}[tb]
\epsscale{0.9}
\plotone{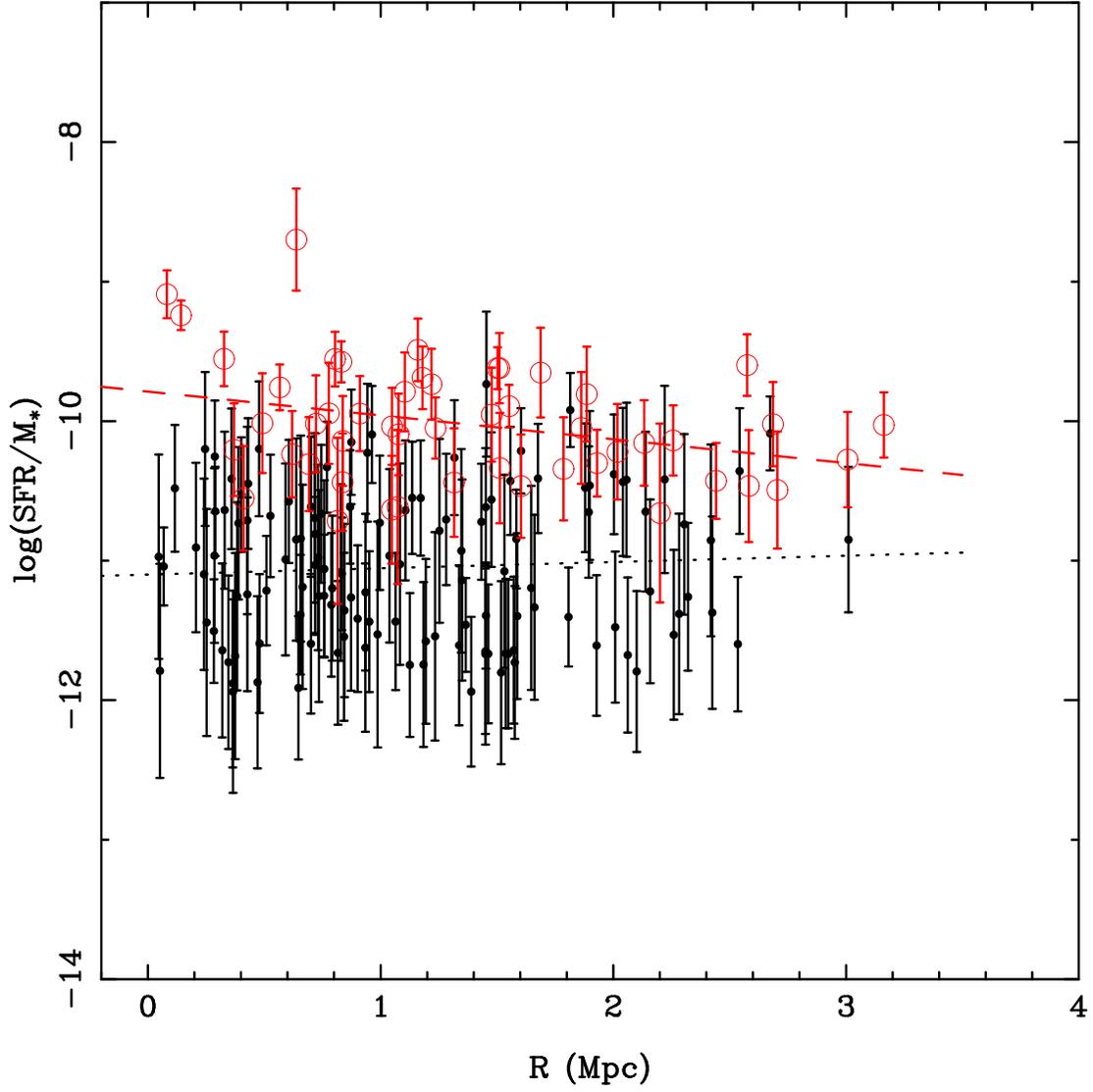}
\caption{The specific SFRs ($SFR/M_*$) as a function of the
distance to the cluster center for the early-type galaxies
({\em solid points}) and the late-type galaxies ({\em
open circles}) in A2255.
\label{sfr_m_r}}
\end{figure}

The star formation properties are fundamentally related with the
local environment. It is interesting to observe how the star formation
activities of the galaxies in a cluster vary with the distances to
the cluster center. Fig.~\ref{sfr_m_r} shows the normalized
SFR by stellar mass as a function of projected distances
($R$) to the cluster center. We adopt the center of A2255 as
($17^h12^m31^s$, $+64^{\circ}05'33''$; J2000.0) which is given by the
NASA/IPAC Extragalactic Database (NED).
At first glance, the striking feature in Fig.~\ref{sfr_m_r} is that the cluster
galaxies with various morphologies behave significantly different
tendencies about how the specific SFRs vary with the locations. For
the early-type galaxies, there is a very weak trend that the inner
galaxies have lower SFR/$M_*$ values, with a linear fitting of 
$\log(SFR/M_*)=0.05(\pm0.07)\,R(\mbox{\rm Mpc}) -11.10(\pm 0.09)$. The
corresponding correlation coefficient is 0.058, and the $rms$
dispersion of $\log(SFR/M_*)$ is 0.522. This trend suggests a slight
suppression of the star formation activity when these cluster galaxies
become massive and concentrated. The inner early-type galaxies might
have higher probability to be near the end point of their star
formation activities when the internal gas reservoir is run out.

As a sharp contrast, the late-type galaxies follow a clear trend that the
inner galaxies have higher rate of mass supplement from new stars (relative
to the assembled stellar mass). The correlation between SFR/$M_*$ and
the projected radial distance $R$ is more significant for the late-type
galaxies. We obtain a linear fit of $\log(SFR/M_*)=-0.17(\pm0.07)
\,R(\mbox{\rm Mpc}) - 9.79(\pm0.12)$. Its correlation coefficient is
$-0.492$, and the $rms$ dispersion in $\rm log(SFR/M_*)$ is 0.392. 
The different behaviors in radial distribution of the 
specific SFRs between the early- and late-type galaxies in A2255 might
indicate that the galaxy-scale turbulence has played different roles on
the star formation process for different types of galaxies,
which lead to a suppression of the star formation activity for E/S0
galaxies and a SFR enhancement for spiral and irregular galaxies.

\section{Discussion}

A2255 is a merging cluster with an intermediate redshift ($z=0.0806$),
which provides an ideal laboratory for studying the SFR variation of
the galaxies within a merging cluster with morphology and projected
radial distance. There are 52 late-type
cluster galaxies with the $r$-band absolute magnitude brighter
than $-20.0$. On the one hand, the rich clusters with smaller
redshifts are expected to have smaller number of blue late-type
member galaxies in cluster cores (Butcher \& Oemler 1984). On the
other hand, for the rich clusters with higher redshifts, a lot of
late-type member galaxies will be fainter than $r=17.77$, and thus
beyond the limit of the SDSS spectroscopy.

According to the hierarchical model of large-scale structure formation
in the picture of cold dark matter, clusters grow in mass by accreting
galaxies from the surrounding field. After the galaxies are involved
into the dense environment (e.g., a cluster of galaxies), the further
star formation activity within the galaxies is expected to be
suppressed by some important processes, such as tidal stripping and
``harassment'' (Moore et al. 1996), ram pressure stripping of the gas
disk (Abadi, Moore, \& Bower 1999; Quilis, Moore, \& Bower 2000), and
removal of gas reservoir surrounding each galaxy (Balogh et al.
1999). These galaxies then evolve to be the early-type galaxies
with red color and little gas. This hierarchical model can be used
to interpret why a higher fraction of the early-type galaxies
appear in nearby galaxy clusters rather than in the high-$z$ clusters
and in the field (Dressler 1980; Whitmore et al. 1993).


The correlation between SFR/$M_*$ and stellar mass
($M_*$) has been found for both the early- and late-type galaxies in
A2255, with a trend that the star formation activity declines as the
assembled stellar mass increases. Brinchmann \& Ellis (2000)
constructed a sample of 321 field galaxies with known morphologies in
a large redshift range of $0<z<1$, and found the same trend for
both E/S0 and spiral disk galaxies. This SFR/$M_*-M_*$ correlation
can be extended to some {\em dwarf} early-type galaxies in the field
(with $M_* <3 \times 10^9$ \Msolar), and the
slope for early-type {\em disk} galaxies ($-1.19$) is much
steeper than that for the E/S0 galaxies in A2255 ($-0.23\pm0.15$).
The late-type galaxies in a cluster seem to have similar behavior
in SFR/$M_*-M_*$ relation as compared with the field spiral and
peculiar galaxies. It should be noted that there are only two massive
late-type galaxies (i.e., $M_*>10^{11}$ \Msolar) survived in A2255,
which might indicate that the massive late-type galaxies have greater
probability to collide with other galaxies as they are falling into
a cluster, and their gas disks could be tidally stripped. Considering
the fact that about half of spiral galaxies in the Virgo cluster show 
the truncated \Ha\ disks (Koopmann et al. 2004), a certain fraction of
the late-type galaxies in A2255 might have also experienced the 
ICM-ISM stripping and strong tidal interaction. As a result, some 
massive spiral galaxies have died out their star formation activities, 
and then evolved into the early-type galaxies.



The correlation between the SFR/$M_*$ and the projected radial
distance $R$ is marginally found for the early-type galaxies in A2255.
This slight trend of inner decrease in their specific SFRs 
can be well interpreted with the hierarchical model: once
the galaxies enter the denser environment (i.e., the core region
of a cluster), their star formation activities have been reduced
by the above-mentioned physical processes, and their appearances
tend to become more concentrated. 
However, some spiral/irregular galaxies in A2255 have survived from
those processes, which means that the overall cluster tidal field
or high speed ``harassment'' interaction between galaxies should have
not stripped their gas disks during the formation of this cluster
(via accreting and merging of subunits). Of great interest is that
these late-type galaxies are found to possess a contrary trend that
the inner late-type galaxies are likely to have more violent star
formation activities. 

Is the morphological dependence of the SFR/$M_*-R$ relation universal
for the galaxies in rich clusters? This tendency has also been found
in some nearby and high-$z$ clusters of galaxies.
Moss \& Whittle (2000) use an objective prism technique to
survey the \Ha\ emission from the late-type galaxies in the regions of
eight nearby Abell clusters ($z<0.024$), and they find that the frequency 
of starburst emission in spiral galaxies increases from the regions
of lower to higher local galaxy density. For the nearby galaxy cluster 
A1367 at $z$=0.0214, a much larger fraction of spiral galaxies are
detected with compact \Ha\ emission in the cluster core region
(Moss, Whittle, \& Pesce 1998). The similar radial distribution
of the SFRs is found for the high-$z$ galaxy cluster CLJ0023+0423B
($z=0.84$): a high fraction of the late-type galaxies (characterized by
the small index given by the Sersic profile fittings) with strong \Ha\
emission are found in the cluster core (Finn, Zaritsky, \& McCarthy
2004). The radial trend of the SFR distribution for the late-type 
galaxies in the Virgo, Coma and A1367 cluster is studied by
Gavazzi et al. (2002), and no clear trend is found for these 
three local clusters. Only the sample of bright Virgo galaxies shows 
an inner decrease in SFR, which is inconsistent with our finding for
the late-type galaxies in A2255. 
  
What physical processes may lead to an enhancement of star formation 
activity for the inner late-type galaxies? This question is
important for understanding the evolution of cluster galaxies.
Lavery \& Henry (1988) first proposed that the star formation can be
triggered by galaxy-galaxy interactions in clusters with intermediate
redshifts, which could be used to explain why the fraction of blue
galaxies in cluster core increases with redshift (Butcher \& Oemler
1984). Numerical simulations by Gnedin (1999) have shown that the
time-varying cluster potential could cause a sequence of strong
tidal shocks on individual galaxies. The shocks probably take place
over a wide region of the cluster, and enhance the galaxy-galaxy
merger rates. The tidal perturbation in a short range may lead
to a triggering of star formation.

Recent observations from the radio (e.g., Feretti et al. 1997,
Miller \& Owen 2003) and
X-ray bands (Burns et al. 1995, Davis \& White 1998) show some
features indicative of an ongoing merger event in A2255. Based on
dynamical analysis of the member galaxies, we also find a direct
evidence showing that A2255 possesses a cluster/group merger (Yuan
et al. 2003). An ordered magnetic field on large scales is
recently detected in A2255 by Govoni et al. (2005), which strongly 
supports the merging scenario. We would like to point out that the 
subcluster mergers might be 
the main mechanism to drive a time-varying cluster potential and
an accelerated rate of galaxy-galaxy merger, which may have
played different roles on the star formation activities for
present-day early- and late-type galaxies in A2255.
It is worthwhile to
carry out the similar studies to check this effect with the large
and complete samples of the galaxies in the clusters at different
stages of dynamical evolution.



\section{Conclusion}

On the basis of the physical parameters derived by Brinchmann et al.
(2004b) for the SDSS galaxies, this paper explores the star formation
properties of the galaxies in A2255, a merging cluster of galaxies
with the intermediate redshift of 0.0806. The main conclusions can be
summarized as follows:

1. The continuum break strength at 4000\AA\ ($D_n$) is a better
morphology indicator for the {\em cluster} galaxies. For selecting the 
early-type galaxies in high-density regions, the criterion 
$D_n>1.6$  is more effective than the concentration parameter 
$R_{p,90}/R_{p,50}$.

2. For the late-type galaxies in A2255, a strong correlation 
between the EW of \Ha\ emission line and the normalized SFR by stellar 
mass is confirmed. This correlation demonstates the reliability of
the SFR and stellar mass estimates by the SDSS spectral data.

3. We have investigated the specific SFRs for the early-
and late-type galaxies in A2255 as a function of stellar mass ($M_*$)
and continuum break strength at 4000 \AA\ ($D_n$). Generally 
speaking, the early-type member galaxies have lower SFRs per unit
stellar mass than the late-types do. Only two massive
late-type galaxies (with $M_*>10^{11}$ \Msolar) are found in A2255.
The late-type galaxies possess tighter correlations of SFR/$M_*$ with
the assembled stellar mass ($M_*$) and the continuum break strength
($D_n$).

4. We have observed the morphological dependence in radial distribution
of the specific SFRs. The E/S0 galaxies are likely to have an inner
decrease in their star formation activities. On the contrary, 
the inner late-type galaxies tend to have more violent star formation 
activities. The merging scenario might be the main mechanism 
to lead to different influences on the star formation activities of
present-day early- and late-type galaxies in A2255.


\acknowledgments

We acknowledge the anonymous referee for his thorough reading of
this paper and invaluable suggestions.
This work is mainly supported by the National Key Base Sciences
Research Foundation under the contract TG1999075402 and is also
supported by the Chinese National Science Foundation
(NSF) under the contract No.10273007. This research has made
use of the NASA/IPAC Extragalactic Database (NED), which is
operated by the Jet Propulsion Laboratory at Caltech, under
contract with the NASA.



\end{document}